\pdfoutput=1
\documentclass[a4paper,11pt]{article}
\usepackage{pos,physics}
\usepackage{graphicx}
\usepackage{subcaption}
\usepackage{multirow, longtable}
\usepackage{bm}
\usepackage{hyperref}
\graphicspath{ {./plots/} }

\title{Nucleon Form Factors from the Feynman-Hellmann Method in Lattice QCD}

\author*[a]{M.~Batelaan}
\author[b]{R.~Horsley}
\author[c]{Y.~Nakamura}
\author[d]{H.~Perlt}
\author[e]{D.~Pleiter}
\author[f]{P.E.L.~Rakow}
\author[g]{G.~Schierholz}
\author[h]{H.~St\"uben}
\author[a]{R.~D.~Young}
\author[a]{J.~M.~Zanotti}

\affiliation[a]{CSSM, Department of Physics, University of Adelaide, Adelaide SA 5005, Australia}
\affiliation[b]{School of Physics and Astronomy, University of Edinburgh, Edinburgh EH9 3FD, United Kingdom}
\affiliation[c]{RIKEN Advanced Institute from Computation Science, Kobe, Hyogo 650-0047, Japan}
\affiliation[d]{Institut f\"ur Theoretische Physik, Universit\"at Leipzig, 04103 Leipzig, Germany}
\affiliation[e]{PDC Center for High Performance Computingl KTH Royal Institute of Technology, SE-100 44 Stockholm, Sweden}
\affiliation[f]{Theoretical Physics Division, Department of Mathematical Physics, University of Liverpool, Liverpool L69 3BX, United Kingdom}
\affiliation[g]{Deutsches Elektronen-Synchrotron DESY, Notkestr. 85, 22607 Hamburg, Germany}
\affiliation[h]{Univerist\"at Hamburg, Regionales Rechenzentrum, 20146 Hamburg, Germany}

\emailAdd{mischa.batelaan@adelaide.edu.au}

\abstract{{\bf QCDSF-UKQCD-CSSM Collaboration}\\ \\ 
 Lattice QCD calculations of the nucleon electromagnetic form factors are of interest at both the high and low momentum transfer regions.
  For high momentum transfers especially there are open questions which require more intense study, such as the potential zero crossing in the proton's electric form factor.
  We will present recent progress from the QCDSF/UKQCD/CSSM collaboration on the calculation of these form factors using the Feynman-Hellmann method in lattice QCD.
  The Feynman-Hellmann method allows for greater control over excited states which we take advantage of by going to high values of the momentum transfer.
  In this proceeding we present results of the form factors up to 6 $\textrm{GeV}^{2}$, using $N_{f}=2+1$ flavour fermions for three different pion masses in the range 310-470 $\textrm{MeV}$.
  The results are extrapolated to the physical pion mass through the use of a flavour breaking expansion.}

\FullConference{%
 The 38th International Symposium on Lattice Field Theory, LATTICE2021
  26th-30th July, 2021
  Zoom/Gather@Massachusetts Institute of Technology
}

\begin{document}
\maketitle

\section{Introduction}
The nucleon electromagnetic form factors are of great interest for furthering the understanding of the internal structure of the nucleon as they describe the distribution of the magnetisation and electric charge inside the nucleon. 
The behaviour of these form factors with respect to $Q^{2}$ near the zero-momentum point determines the charge and magnetisation radius of the nucleon. This behaviour has been studied extensively using lattice QCD \cite{Boinepalli:2006xd,Green:2014xba,Alexandrou:2006ru} and through experiments \cite{Perdrisat:2006hj,Arrington:2006zm,Arrington:2011kb}. The large momentum behaviour is also of interest as the currently available experimental results show a decrease in the ratio of $G_{E}/G_{M}$ at large $Q^{2}$ \cite{Punjabi:2005wq, Puckett:2010ac} raising the question of whether \(G_{E}(Q^{2})\) crosses zero at some large value of \(Q^{2}\).
The matrix element of the electromagnetic current, \(j_{\mu}\), can be written down in terms of the Dirac ($F_{1}$) and Pauli ($F_{2}$) form factors, in Euclidean space this matrix element is defined as
\begin{align}
\label{eq:1}
  &\mel{N(p',s')}{j_{\mu}(q)}{N(p,s)} = \bar{u}_{N}(p',s')\left[ \gamma_{\mu}F_{1}(Q^{2}) + \sigma_{\mu\nu}\frac{q_{\nu}}{2M_{N}}F_{2}(Q^{2})\right] u_{N}(p,s),
\end{align}
where \(q=p'-p = - Q^{2}\), \(\sigma_{\mu\nu} = \frac{i}{2}[\gamma_{\mu},\gamma_{\nu}]\) and \(u_{N}(p,s)\) is a Dirac spinor with momentum \(p\) and spin polarization \(s\).
The proton electromagnetic form factors can then be calculated by using the above equation with the following electromagnetic current
\begin{equation}
\label{eq:8}
j_{\mu} = \frac{2}{3}\bar{u}\gamma_{\mu}u - \frac{1}{3}\bar{d}\gamma_{\mu}d,
\end{equation}
neglecting contributions from non-valence quarks.
The Sachs electromagnetic form factors will be used in this proceedings, they are written as combinations of the Dirac and Pauli form factors
\begin{align}
\label{eq:2}
  G_{E}(Q^{2}) &\equiv F_{1}(Q^{2}) - \frac{Q^{2}}{4m^{2}}F_{2}(Q^{2}), \\
  G_{M}(Q^{2}) &\equiv F_{1}(Q^{2}) + F_{2}(Q^{2}).
\end{align}
The Feynman-Hellmann method for lattice QCD will be used here to calculate these form factors at high momentum transfers. The Feynman-Hellmann method has seen success in calculations of a number of different observables in lattice QCD \cite{QCDSF:2012mkm,CSSM:2014uyt, Chambers:2015bka, Hannaford-Gunn:2021mrl}.

\section{Lattice Methodology}
The gauge fields used here are generated with $N_{f}=2+1$ flavours of $\mathcal{O}(a)$-improved clover fermions and a tree-level Symanzik-improved gluon action. The volume of the lattice is $L^{3}\times T=32^{3}\times 64$ with a lattice spacing of $a=0.074(2) \textrm{ fm}$, the scale was set using singlet quantities as detailed in refs. \cite{Bornyakov:2015eaa,Bietenholz:2010jr}. The selected hopping parameters give pion masses in the range $310-470 \textrm{ MeV}$. We use one ensemble at the $SU(3)_{\textrm{flavour}}$-symmetric point and two ensembles away from this point, where the singlet quark mass $\bar{m} = \frac{1}{3}(m_{u}+m_{d}+m_{s})$ is kept constant.

\subsection{Two-point Correlation functions}
The momentum projected two-point functions which are calculated on the lattice ensembles are defined by
\begin{align}
\label{eq:13}
  &G_{\chi\bar{\chi}}(\vb*{p};t,t') = \sum_{\vb*{x}}e^{-i\vb*{p}\cdot \vb*{x}} \Gamma^{\textrm{proj.}}_{\alpha\beta}\expval{\chi_{\alpha}(t,\vb*{x}) \bar{\chi}_{\beta}(0, \vb*{0})}{\Omega},
\end{align}
where the source is located at the origin and  $x$ is the sink location. $\Gamma^{\textrm{proj.}}$ is the projection matrix and the interpolating operator for the proton is defined as
\begin{equation}
\label{eq:14}
\chi_{\alpha}(x) = \epsilon^{abc}[u(x)]^{a}_{\alpha}\left( [u(x)]^{b}_{\beta}[C\gamma_{5}]_{\beta\gamma}[d(x)]^{c}_{\gamma} \right),
\end{equation}
where $C$ is the charge conjugation matrix. The interpolating operators will create and annihilate all states with the quantum numbers of the nucleon.
To improve the overlap of the operators with the ground state nucleon, we apply a gauge-invariant Jacobi smearing to the operators at the source and the sink \cite{UKQCD:1993gym}.

\subsection{Feynman-Hellmann Method}
The Feynman-Hellmann theorem in quantum mechanics relates the derivative of the energy to the expectation value of the derivative of the Hamiltonian.
\begin{equation}
\label{eq:4}
\frac{\partial E_{n}(\lambda)}{\partial \lambda} = \ev{\frac{\partial H(\lambda)}{\partial \lambda}}{\psi_{n}(\lambda)}
\end{equation}
This theorem can be extended to lattice QCD by making a modification to the Lagrangian.
The modification we will use here takes the following form
\begin{equation}
\label{eq:3}
\mathcal{L}(x)  \to \mathcal{L}(x) + \lambda_{\mu} \left(e^{i\vb*{q}\cdot \vb*{x}}+e^{-i\vb*{q}\cdot \vb*{x}}\right) \bar{q}(x) \gamma_{\mu} q(x)
\end{equation}
where the $\mu$ is not summed over. If the modification in Eq. \ref{eq:3} uses the temporal current ($\gamma_{4}$), this will lead to a determination of the electric form factor while choosing a component of the spatial current ($\gamma_{i}$) will lead to the magnetic form factor.
In the Breit frame ($\bm{p}' = -\bm{p}$), and for a set of states which diagonalise the derivative of the Hamiltonian \cite{Chambers2017}, the Feynman-Hellmann theorem relates the resulting shift in energy to the electromagnetic form factors
\begin{align}
\label{eq:21}
  \frac{\partial E}{\partial \lambda_{4}}\bigg|_{\lambda=0} &= \frac{M_{N}}{E_{N}}G_{E}(Q^{2})\\
  \label{eq:22}
  \frac{\partial E}{\partial \lambda_{i}}\bigg|_{\lambda=0} &= \frac{[\bm{\hat{e}}\times \bm{q}]_{i}}{2E_{N}}G_{M}(Q^{2}),
\end{align}
where $\bm{\hat{e}}$ is the unit vector in the direction of the spin polarization projection. For the calculations presented here it is fixed to the third spatial direction, \(\hat{\bm{e}}=(0,0,1)\).
To calculate the magnetic form factor we will use the spatial current $\gamma_{2}$ and set the first component of $\bm{q}$ to be non-zero such that the cross product in Eq. \ref{eq:22} does not vanish.

To extract the energy shift arising from the Feynman-Hellmann modification, we construct ratios of correlators with and without the modification.
These ratios will give an effective energy shift which can be compared with different fit functions.
The Feynman-Hellmann method as described above will only give a linear energy shift in the Breit frame, which therefore restricts the momentum transfers which can be accessed through this method. However it also means that for every value of the momentum transfers $Q^{2}$, the state momentum $|\bm{p}|$ is minimized which reduces the signal-to-noise ratio of the correlator. Table \ref{tab:2} shows the values of the momentum transfer which are used here.

\begin{table*}[h]
  \caption{The momentum transfer values which are considered here. The choices are restricted to the Breit frame where $\bm{p'} = -\bm{p}$.}
  \label{tab:2}
  \centering
  \begin{tabular}{c c c}
    \hline\hline
    $\bm{q}(L/2\pi)$ & $\bm{p}(L/2\pi)$ & $Q^{2}(L/2\pi)^{2}$ \\
    \hline
    (0,0,0) & (0,0,0) & 0\\
    (2,0,0) & $\pm$(1,0,0) & 4\\
    (2,2,2) & $\pm$(1,1,1) & 12\\
    (4,2,0) & $\pm$(2,1,0) & 20\\
  \end{tabular}
\end{table*}

The size of $\lambda$ needs to be small enough such that the energy shift remains in the linear regime, but large enough such that it remains possible to extract a non-zero signal. We use a value of $\lambda=10^{-4}$ for this analysis and we use a second value of $\lambda=2\times10^{-4}$ for selected momenta to confirm that the $\lambda$-dependence is linear.
We re-analyse data from Ref. \cite{Chambers2017} to quantify the excited state contamination and to add another point to the flavour breaking expansion.
For these results it has been shown that the chosen values of $\lambda$ are in the linear regime \cite{Chambers2017}. For the other ensembles used here, we use the same value for $\lambda$.

\subsection{Contamination by Excited States}
The main difficulty in calculating form factors  at high momenta is the deterioration of the signal-to-noise ratio at early time slices. This problem with nucleon correlation functions at high momenta has been known for a long time \cite{Lepage:1989hd,Parisi:1983ae}.

We will consider momentum projections of the two-point correlators up to $6 \textrm{ GeV}^{2}$. In order to extract energies from these correlators in a consistent manner over this large range of momenta, we will use two fitting functions and apply a weighted averaging method \cite{NPLQCD:2020ozd} to the results.
The spectral decomposition of the two-point functions on the lattice reduces to an infinite sum of exponentials. In order to accurately extract the energy of the nucleon ground state we will consider the first two terms of this sum. The second term will then encompass all the excited states beyond the ground state.
These two models will be used to fit the data over a range of fit windows,
\begin{align}
\label{eq:15}
  G_{a}(\vb*{p};t) &= A_{0}e^{-E_{0}t}, \\
  \label{eq:16}
G_{b}(\vb*{p};t) &= A_{0}e^{-E_{0}t} + A_{1}e^{-E_{1}t}.
\end{align}
The shift in the energy of the correlation functions is required to calculate the form factors. To make use of the correlations in the data we construct ratios of the perturbed and unperturbed correlators, this will give a cleaner signal than fitting the two correlators separately.
To make full use of the correlations we use time-reversed, both parity projections and equivalent momentum projections to construct the ratios. The first ratio will give the electric form factor from the temporal component of the current
\begin{equation}
\label{eq:8}
R_{E,p}(\bm{p},t) = \left|\frac{\bar{G}^{+}(\bm{p},\lambda,t)\bar{G}^{-}(\bm{p},0,-t)}{\bar{G}^{+}(\bm{p},0,t)\bar{G}^{-}(\bm{p},\lambda,-t)}\right|^{\frac{1}{2}},
\end{equation}
and the second gives the magnetic form factor from the spatial component of the current
\begin{align}
\label{eq:5}
  R_{M,p}(\bm{p},t) = \left|\frac{G^{\pm}_{\uparrow}(\bm{p},\lambda,t)G^{\pm}_{\downarrow}(\bm{p},0,t)G^{\pm}_{\uparrow}(-\bm{p},\lambda,t)G^{\pm}_{\downarrow}(-\bm{p},0,t)}{G^{\pm}_{\uparrow}(\bm{p},0,t)G^{\pm}_{\downarrow}(\bm{p},\lambda,t)G^{\pm}_{\uparrow}(-\bm{p},0,t)G^{\pm}_{\downarrow}(-\bm{p},\lambda,t)}\right|^{\frac{1}{4}}.
\end{align}
Where we have defined $\bar{G}(\bm{p},\lambda,t) = \frac{1}{2}[G(+\bm{p},\lambda,t) + G(-\bm{p},\lambda,t)]$, the \(+, -\) superscripts indicate the positive and negative parity projections respectively ($\Gamma_{\pm} = \frac{1}{2}(1\pm \gamma_{4})$) and we define $G^{\pm}(\bm{p},\lambda,t) = \frac{1}{2}[G^{+}(\bm{p},\lambda,t)+G^{-}(\bm{p},\lambda,-t)]$. 

These ratios we will then fit with two functions again, one which only includes the ground state and another which includes both the ground state and an excited state.
The first fit function will only be valid in the large Euclidean time limit, when the ground state has saturated.
\begin{equation}
\label{eq:11}
R_{a}(\bm{p},t) \xrightarrow[]{t \gg 0} A(\lambda) e^{-\Delta E(\lambda) t}
\end{equation}
The second fit function will use the energies and amplitudes of the two states of the unperturbed correlators defined in Eq. \ref{eq:16}, and explicitly fits to the energy shift for both the ground state and the excited states,
\begin{equation}
\label{eq:12}
  R_{b}(\bm{p},t) =\frac{(A_{0}+\Delta A_{0})e^{-(E_{0}+\Delta E_{0})t} + (A_{1}+\Delta A_{1})e^{-(E_{1}+\Delta E_{1})t}}{(A_{0}-\Delta A_{0})e^{-(E_{0}-\Delta E_{0})t} + (A_{1}-\Delta A_{1})e^{-(E_{1}-\Delta E_{1})t}}.
\end{equation}
This fit function will describe the data better at smaller Euclidean time which will allow quantitative monitoring of the contamination from excited states.

To reduce the influence of the choice of fit window we apply a weighted averaging method to the fit results. This method will take results from different fit windows and from the two fit functions and assign to each result a weight. The weighted averaging method we use is presented in \cite{NPLQCD:2020ozd} and \cite{Rinaldi:2019thf}, it uses a modified inverse-variance weighting where the weight depends on the $\chi^{2}_{\textrm{dof}}$ value and the uncertainty of the fit parameters.

\begin{equation}
\label{eq:5}
\tilde{w}^{f} = \frac{p_{f}\left(\delta E_{0}^{f}\right)^{-2} }{\sum_{f'=1}^{N}p_{f'}\left(\delta E_{0}^{f'}\right)^{-2}},
\end{equation}
Where \(p_{f}\) is the p-value of the fit \(f\) and \(\delta E_{0}^{f}\) is the uncertainty in the value of the ground state energy of fit \(f\). This weighting method allows for the results of both fit functions to be used in a consistent manner across a large range of \(Q^{2}\) values.

\begin{figure}[h]
  \begin{subfigure}{.5\textwidth}
  \centering
  \includegraphics[width=\linewidth]{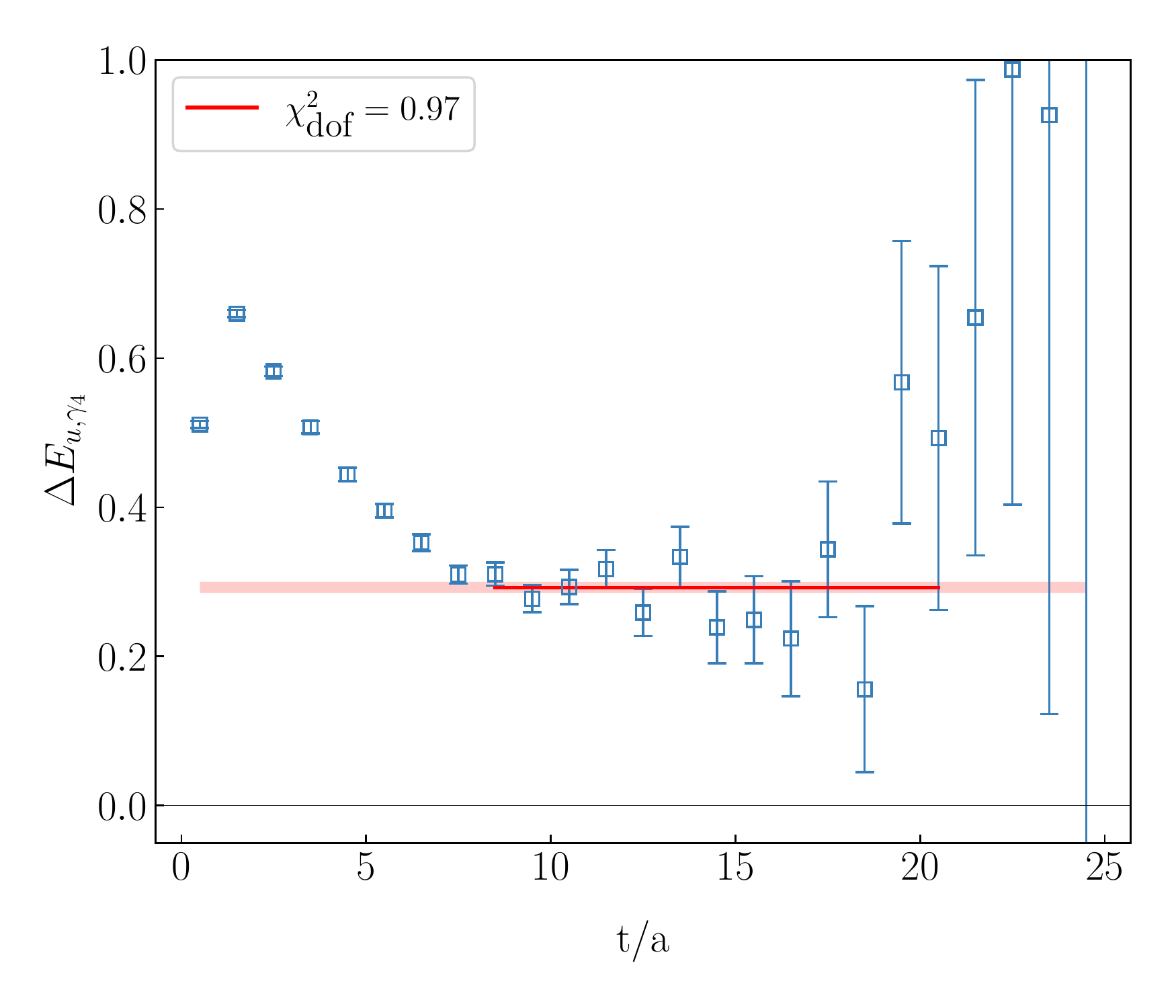}
  \end{subfigure}
  \begin{subfigure}{.5\textwidth}
    \centering
    \includegraphics[width=\linewidth]{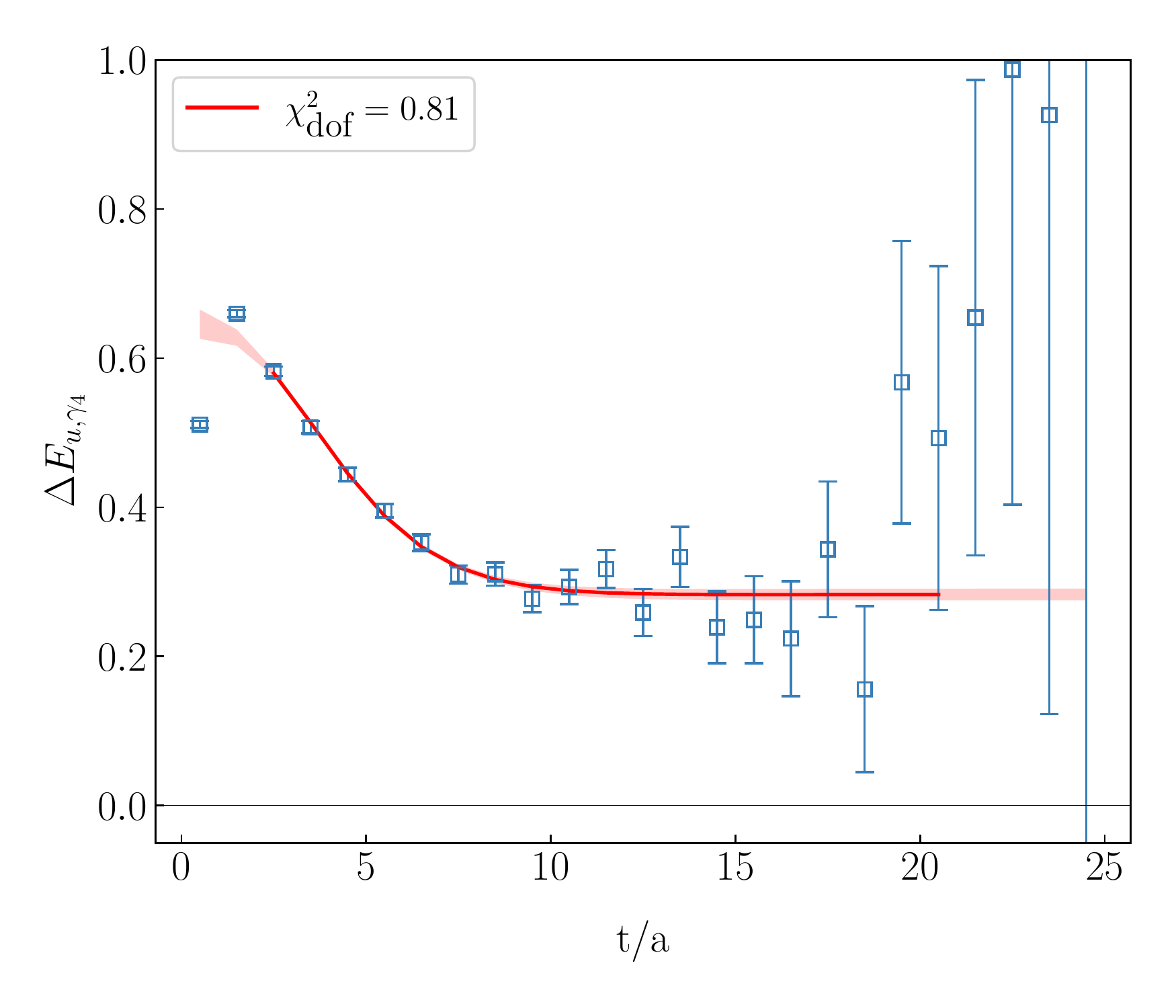}
  \end{subfigure}
  \caption{The effective energy of the ratio of correlators for the up quark in the proton. The effective energy of the one-exponential fit and the two-exponential fit to the ratio are also shown with their respective $\chi^{2}_{\textrm{dof}}$ values.}
  \label{fig:one-exp}
\end{figure}

\begin{figure}[h]
  \begin{subfigure}{.5\textwidth}
  \centering
  \includegraphics[width=\linewidth]{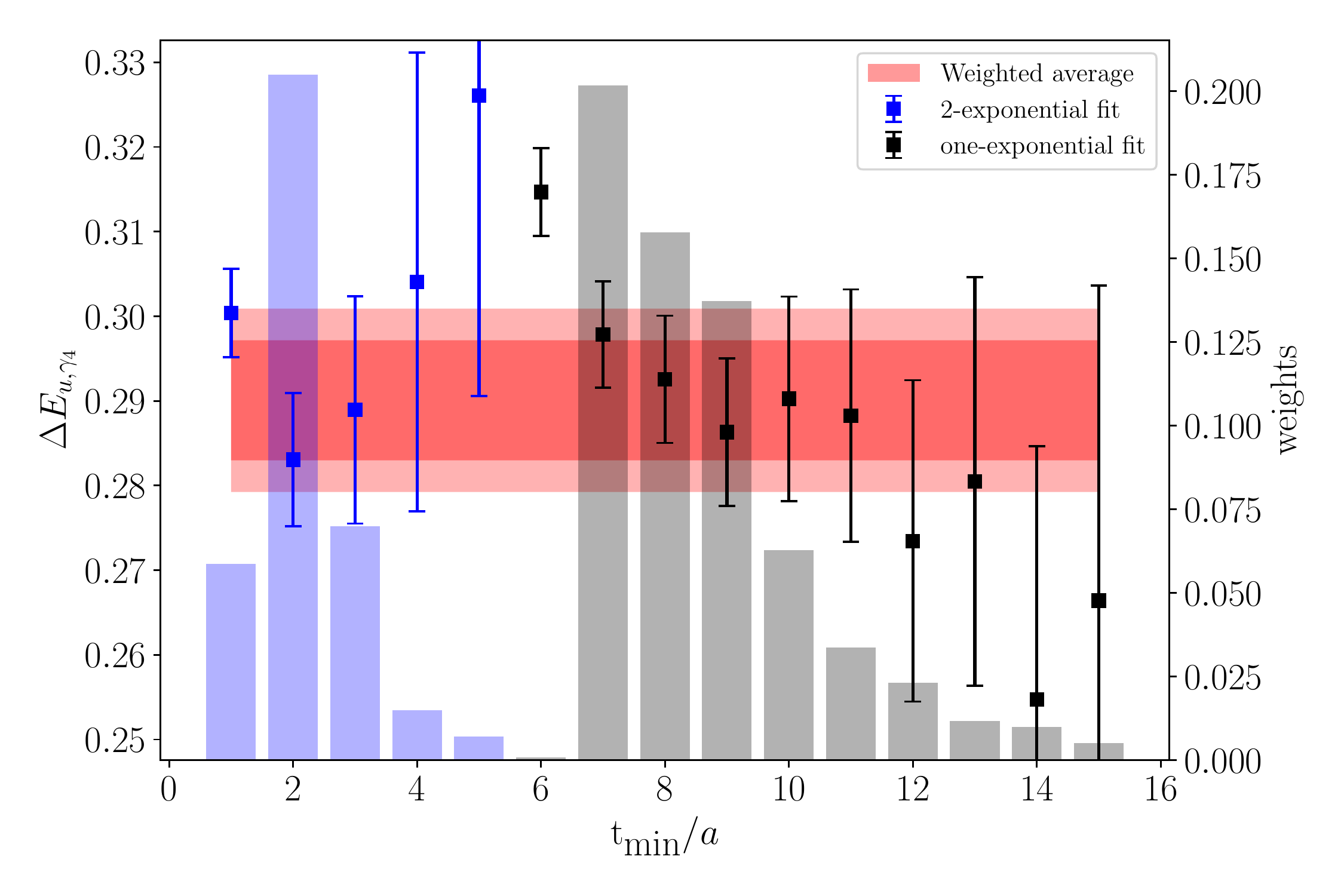}
  \end{subfigure}
  \begin{subfigure}{.5\textwidth}
  \centering
  \includegraphics[width=\linewidth]{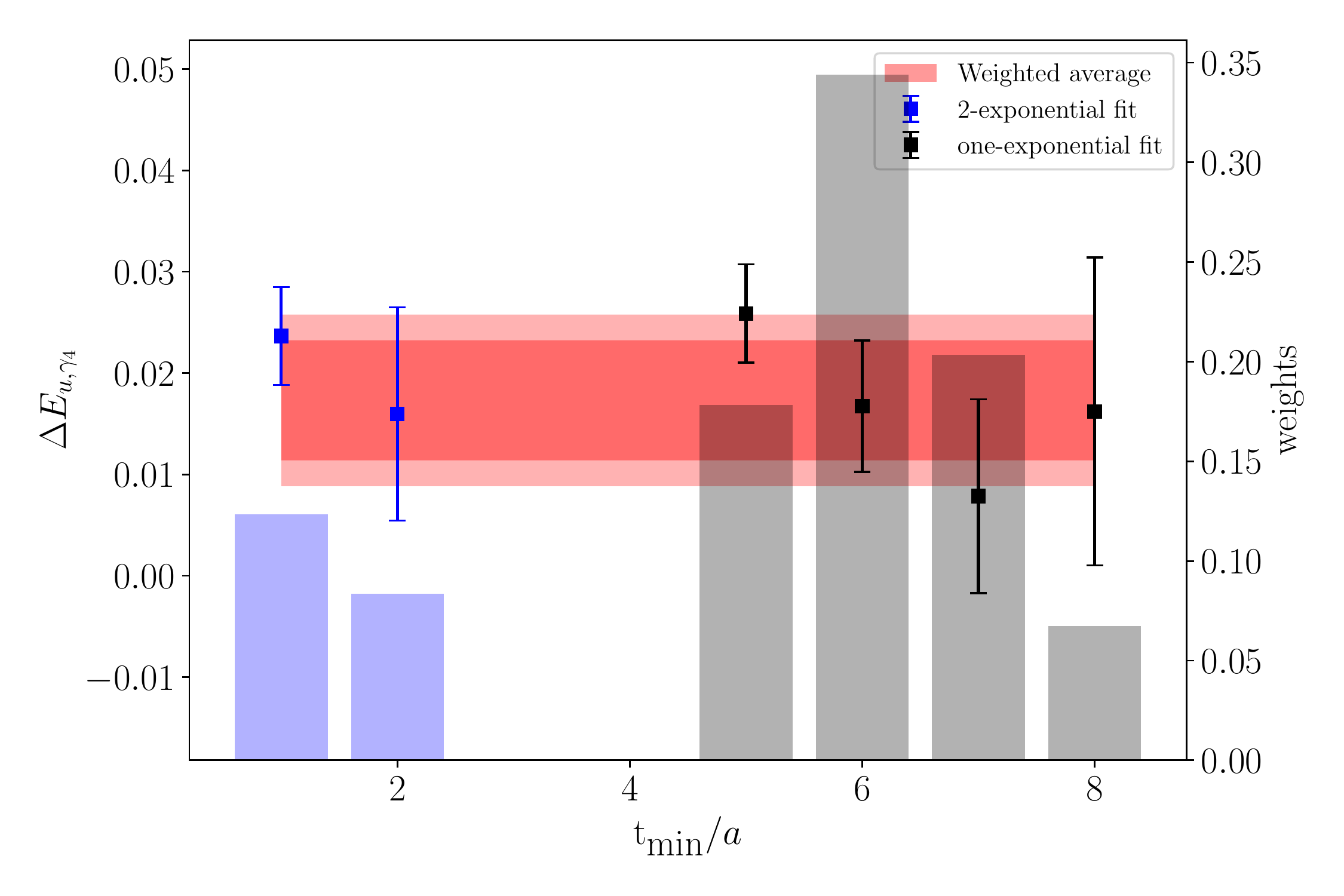}
  \end{subfigure}
  \caption{The energy shift extracted from the fit to the ratio with the two-exponential function (blue points) and the one-exponential function (black points). The bar graph shows the weight of each fit result for the value of $t_{\textrm{min}}$ where the blue bars correspond to the two-exponential fit and the black bars to the one-exponential fit. The red band is the weighted average value, where the inner band shows the statistical uncertainty and the outer band shows the total uncertainty, this includes the statistical uncertainty and the systematic uncertainty from the spread between the included fit results. The left hand plot shows the energy shift for the lattice momentum of \(\bm{q}=\frac{2\pi}{L}(2,0,0)\), the right hand plot shows the energy shift for the lattice momentum of \(\bm{q}=\frac{2\pi}{L}(4,2,2)\).}
  \label{fig:weighted_avg}
\end{figure}

Figure \ref{fig:one-exp} shows the results of both fitting functions on the effective energy shift and figure \ref{fig:weighted_avg} shows how the energy shift from the fit changes under variations in $t_{min}$. Figure \ref {fig:weighted_avg} also shows a bar graph for the weights assigned to each fit value, this shows that each function has a $t_{min}$ value for which the weights peak. Including both fit functions in the weighted average gives a more consistent method for extracting the energy shift over the range of momenta.

\section{Flavour Breaking Expansion}
\begin{figure}
  \centering
  \includegraphics[width=0.7\linewidth]{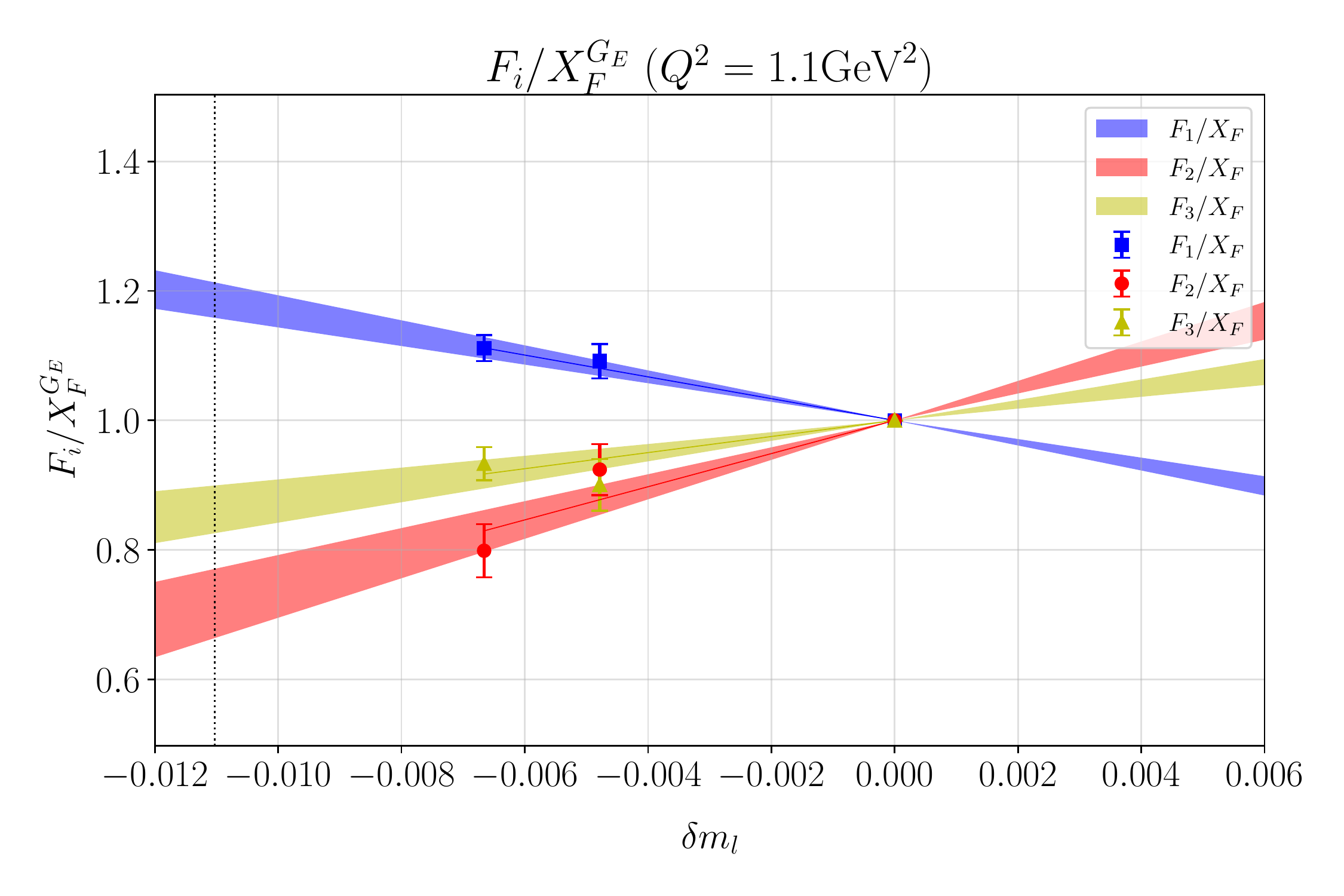}
  \caption{The $F_{i}$ quantities for the electric form factor $G_{E}$ at $Q^{2}=1.1\textrm{ GeV}^{2}$ plotted against the variation in the quark mass. The dotted line represents the physical point. This is from the flavour diagonal form factors of the N, $\Xi$ and $\Sigma$ baryons.}
  \label{fig:fanplot}
\end{figure}

\begin{figure}
  \begin{subfigure}{.5\textwidth}
  \centering
  \includegraphics[width=\linewidth]{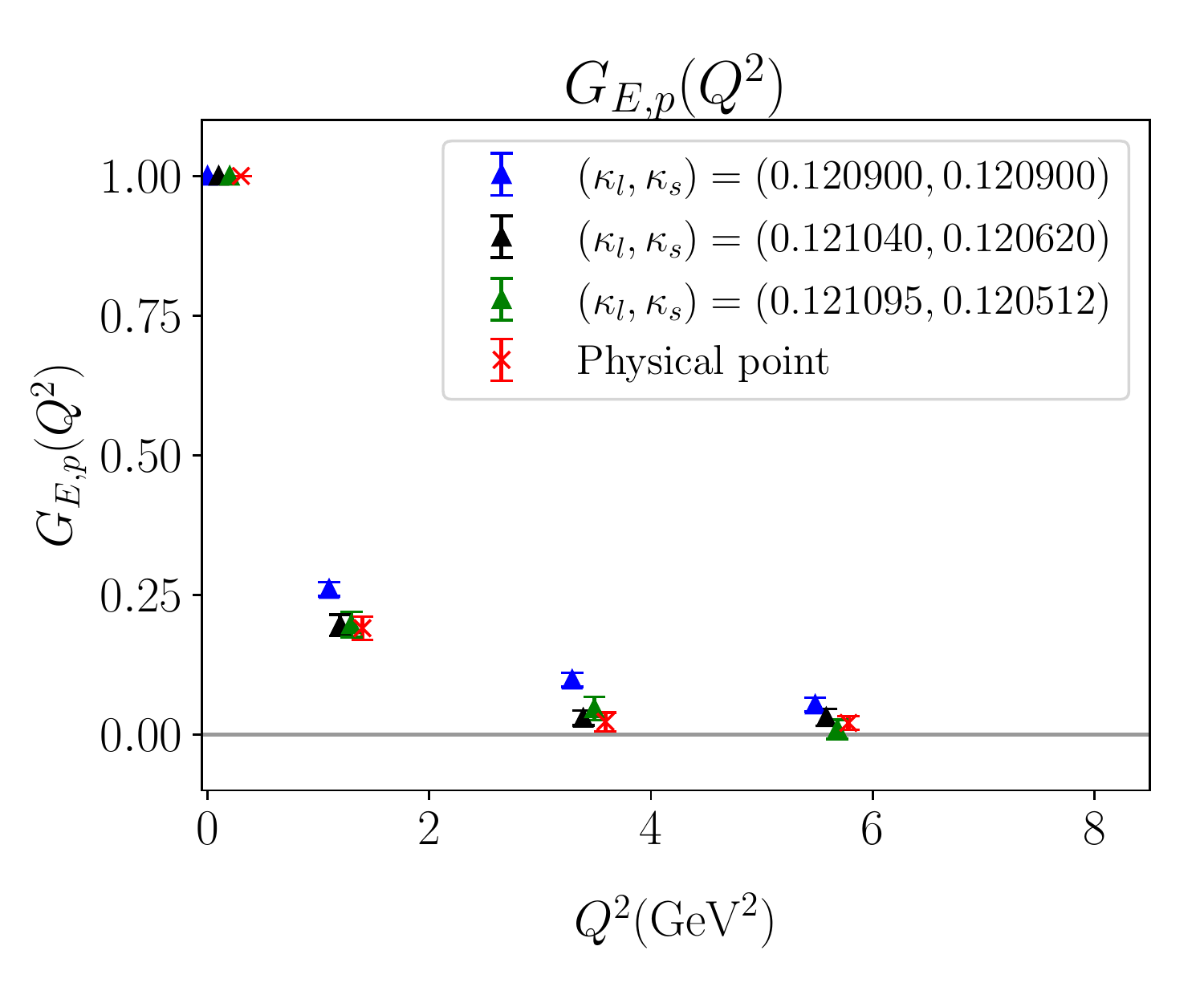}
  \end{subfigure}
  \begin{subfigure}{.5\textwidth}
    \centering
    \includegraphics[width=\linewidth]{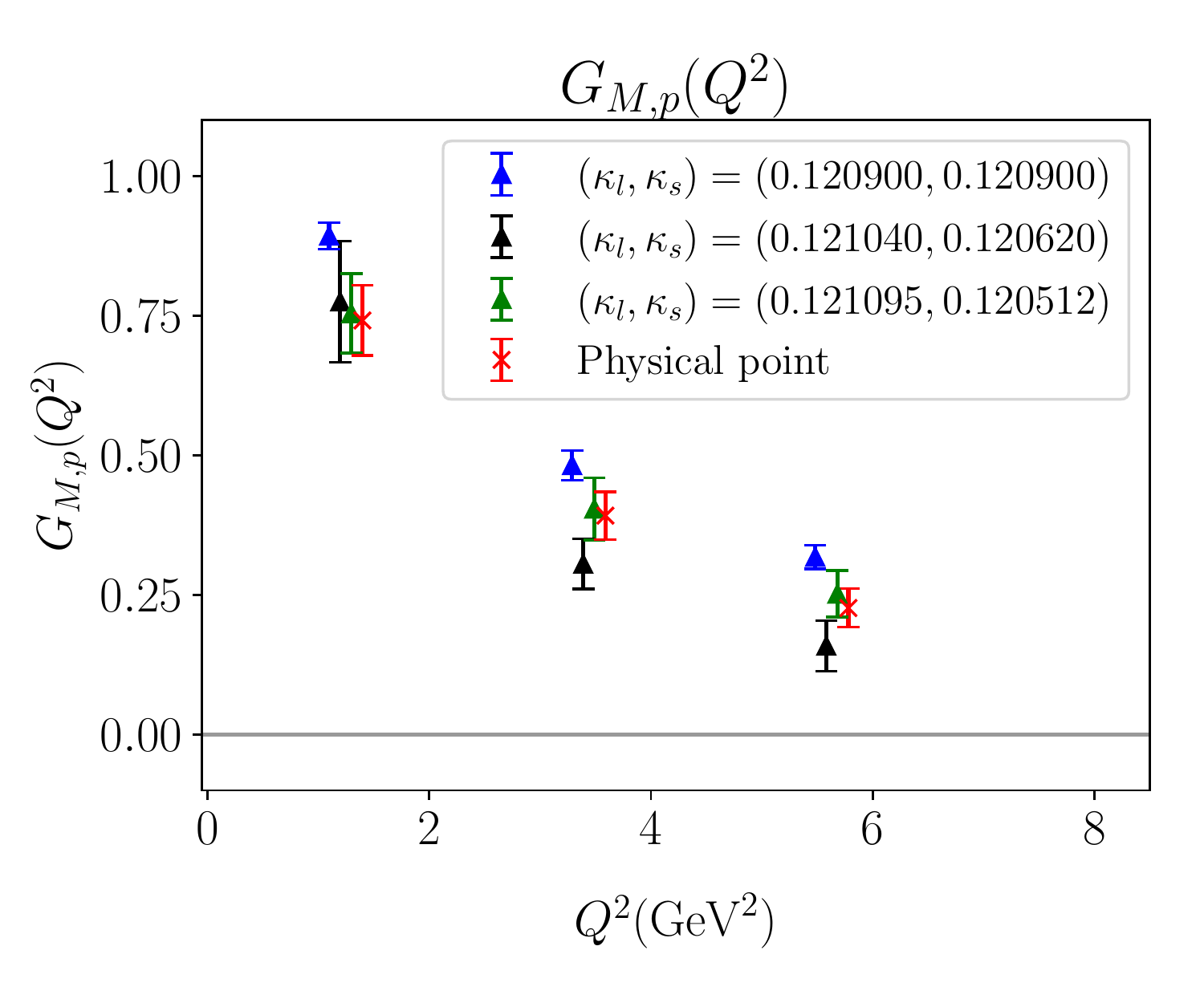}
  \end{subfigure}
  \caption{The electric and magnetic Sachs form factor results on the three lattice ensembles together with the result of the extrapolation to the physical pion mass using the flavour breaking expansion.}
  \label{fig:EMFF}
\end{figure}

Now that we have established a robust method for determining the energy shifts at finite \(\lambda\), these energy shifts are then used to determine the electromagnetic form factors using equations \ref{eq:21} and \ref{eq:22} for each octet baryon.
To extrapolate the form factor results to the physical quark masses we use a flavour breaking expansion which has been detailed in Refs. \cite{Bietenholz:2011qq, Bickerton:2019nyz}. The key element of this expansion is that the average quark mass stays constant,
\begin{equation}
\label{eq:6}
\bar{m} \equiv \frac{1}{3}\left(m_{u} + m_{d}+m_{s}\right).
\end{equation}
Since we are working with $N_{f} = 2+1$ quark flavours, the up-quark and the down-quark have the same mass ($m_{l}$) and hence the expansion will be done in the parameter $\delta m_{l}$ which describes the distance from the SU(3) flavour symmetric point,
\begin{equation}
\label{eq:7}
\delta m_{l} \equiv m_{l} - \bar{m}.
\end{equation}
For the flavour breaking expansion, we calculate seven $D_{i}$ values and five $F_{i}$ values \cite{Bickerton:2019nyz} from the form factors of the various octet baryons. These quantities are constructed such that they have the same value at the SU(3) flavour symmetric point but their values diverge at non-zero $\delta m_{l}$.
We also construct an average $D$ value $X_{D}$ for which the $\delta m_{l}$ component cancels out. The same is done for the $F$ quantities, these will be used to normalise the $D_{i}$ and $F_{i}$ quantities, see \cite{Bickerton:2019nyz} for more details. Figure \ref{fig:fanplot} shows an example of the flavour breaking expansion. The $F_{i}$ can be seen branching out at non-zero $\delta m_{l}$, the vertical dotted line on the plot represents the physical quark masses. We fit a linear function to the results and use this to extrapolate the $F_{i}$ and $D_{i}$ quantities to the physical quark masses. From there we can reconstruct the form factors at the physical point for each value of \(Q^{2}\). Figure \ref{fig:EMFF} shows the results of the electric and magnetic form factors of the proton on the three ensembles as well as the extrapolated values.

\section{Conclusion and Outlook}
The calculation of the electromagnetic form factors with the Feynman-Hellmann method has been made more robust by the explicit inclusion of excited state effects in the fitting function as well as the application of a weighted averaging method to quantify systematic uncertainties in the choice of fit. By including multiple pion masses in the calculation we have been able to produce an extrapolation of these values to the physical point. To improve the form factor results at high momentum transfer, even further, we will expand this analysis to include more lattice spacings, volumes and quark masses in a future publication, this will allow us better control the systematic uncertainties in the results at large \(Q^{2}\).

\section{Acknowledgements}
The configurations used were generated by the the BQCD lattice QCD program \cite{Nakamura:2010qh,Kamleh:2017mpg}. The Chroma Software library \cite{Edwards:2004sx} was used for the data analysis on the DiRAC BlueGeneQ and Extreme Scaling (EPCC, Edinburgh, UK) and Data Intensive (Cambridge, UK) services. The GCS supercomputers JUQUEEEN and JUWELS (NIC, J\"ulich, Germany) and resources provided by HLRN (North-German Supercomputer Alliance), the NCI National Facility in Canberra Australia (supported by the Australian Commonwealth Government) and the Phoenix HPC service (University of Adelaide).
RH is supported by STFC through grant ST/P000630/1. HP is supported by DFG Grant No. PE 2792/2-1. PELR is supported in part by the STFC under contract ST/G00062X/1. GS is supported by DFG Grant No. SCHI 179/8-1. RDY and JMZ are supported by the Australian Research Council grant DP190100297.

\bibliography{Library}

\end{document}